\documentclass[amsmath, amssymb,twocolumn, showpacs]{revtex4-1}

\usepackage{graphicx}
\usepackage{dcolumn}
\usepackage{bm}
\usepackage{color}
\usepackage{latexsym}

\begin{document}

\title{Comparison of splashing in high and low viscosity liquids }

\author{Cacey S. Stevens, Andrzej Latka, and Sidney R. Nagel}

\affiliation{The James Franck Institute and Department of Physics,
The University of Chicago, Chicago, Illinois 60637, USA}

\begin{abstract}
We explore the evolution of a splash when a liquid drop impacts a smooth, dry surface.
There are two splashing regimes that occur when the liquid viscosity is varied, as is evidenced by its dependence on ambient gas pressure.
A high-viscosity drop splashes by emitting a thin sheet of liquid from a spreading liquid lamella long after the drop has first contacted the solid.
Likewise, we find that there is also a delay in the ejection of a thin sheet when a low-viscosity drop splashes.  
We show how the ejection time of the thin sheet depends on liquid viscosity and ambient gas pressure. 
\end{abstract}


\maketitle

The discovery by Xu \textit{et al.} \cite{Xu}, that the splash of a liquid drop hitting a smooth dry surface is suppressed by lowering the ambient air pressure, has galvanized research on gas-liquid interactions during impact.
However, despite numerous experimental  \cite{Xu2,Liu,Driscoll,Driscoll2,kolinski2012skating,de2012dynamics,latka2012creation}, theoretical \cite{Mani,Mandre}, and numerical \cite{Liu,duchemin2011curvature,bang2011assessment} efforts, the mechanism by which air causes a drop to splash remains unresolved. 

\begin{figure}[h] 
\begin{center}
\includegraphics[width=3.1 in]{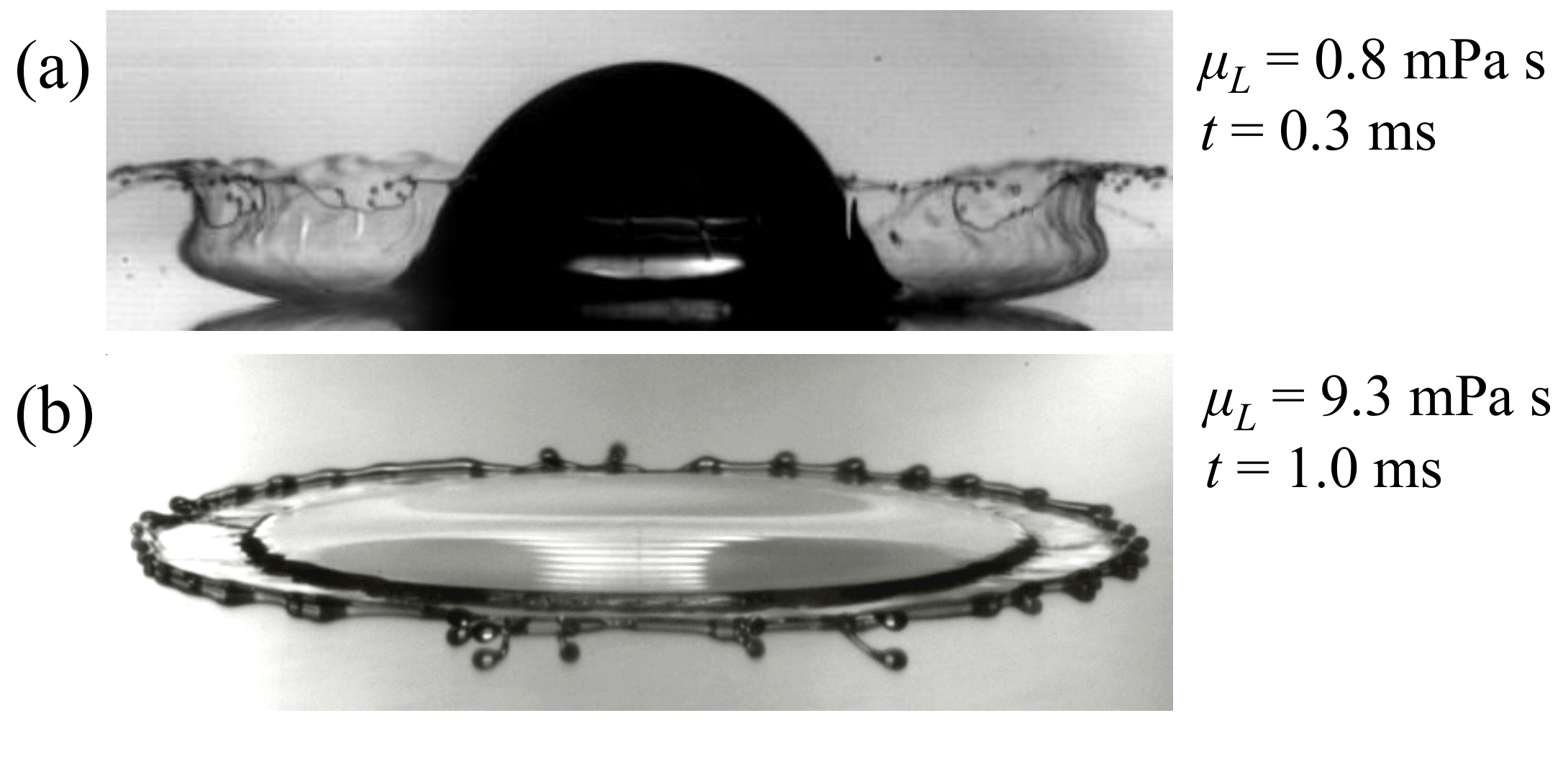}
\caption{Images of 3.2 mm diameter silicone oil drops in the low- and high-viscosity regimes at atmospheric pressure after impacting a smooth surface at $u_0$=4.0 m/s. 
(a) For a 0.8 mPa$\thinspace$s drop, a corona emerges at a large angle from the surface almost immediately after impact.
(b) The splash of a 9.3 mPa$\thinspace$s drop occurs through the ejection of a thin liquid sheet from a thicker lamella. }
\label{images}
\end{center}
\end{figure}

\begin{figure}[h] 
\begin{center}
\includegraphics[width=3.1 in]{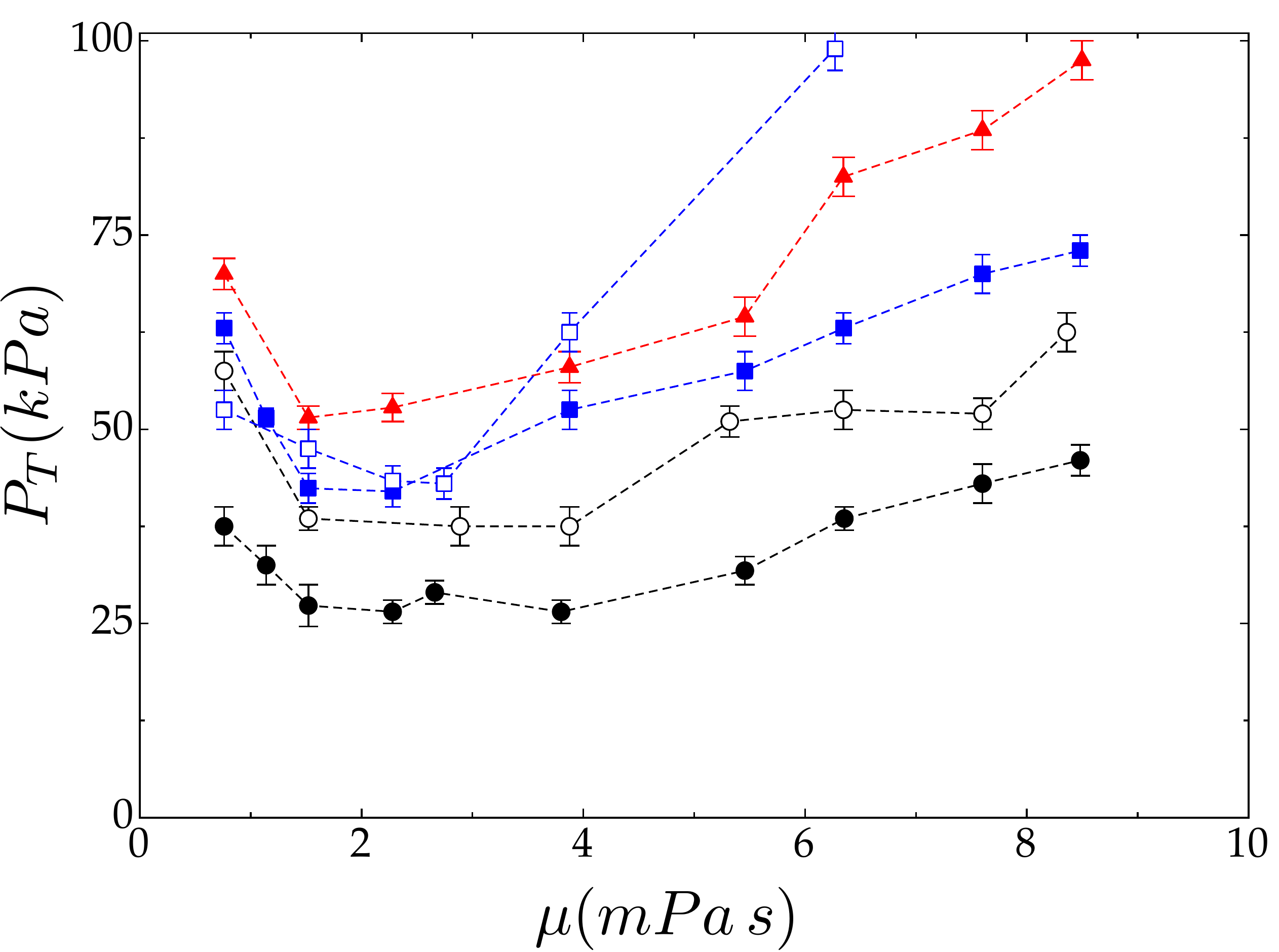}
\caption{Splashing threshold pressure $P_T$ versus liquid viscosity $\mu$ for silicone oil drops. The impact velocity was fixed at 
$u_0$=$4.0\pm0.1$ m/s with drop diameter $D$= $1.1\pm0.1$ mm ({\color{red}$\blacktriangle$}), 
$1.8\pm0.1$ mm ({\color{blue}$\blacksquare$}) or $3.2\pm0.1$ mm ({\color{black}$\bullet$}), fixed at $u_0$ = $2.0\pm0.1$ m/s with D= $3.2\pm0.1$ mm ({\color{black}$\circ$}) or fixed at $u_0$ = $3.0\pm0.1$ m/s with D= $1.8\pm0.1$ mm ({\color{blue}$\Box$}).}
\label{Pt}
\end{center}
\end{figure}

The situation is made more complicated, by the influence of liquid viscosity $\mu$ on the interplay of gas and liquid. 
At low viscosities, a beautiful crown-shaped corona emerges almost immediately after impact as shown in Fig. \ref{images}(a) \cite{Xu,Xu2}.
However, a small increase in viscosity reveals a splash with a strikingly different appearance,  that evolves much more slowly (Fig. \ref{images}(b)). 
This higher-$\mu$ drop first contacts the surface and then spreads smoothly as a thick lamellar sheet.
From this lamella, a thinner sheet of liquid is subsequently ejected almost parallel to the substrate. It is the thin sheet that eventually breaks apart to form the splash \cite{Driscoll}. The existence of two distinct splashing regimes is made manifest in the non-monotonic dependence of the threshold pressure, $P_T$, which is the ambient gas pressure above which splashing occurs, on the viscosity \cite{Xu2}. As shown in Fig. \ref{Pt}, $P_T$ decreases with increasing viscosity at low-$\mu$, while the trend is reversed at higher $\mu$. 

These differences have been taken to suggest that distinct mechanisms might underlie the two types of splash. Indeed, theories for low-$\mu$ splashes have been proposed that do not take into account any spreading of a liquid film on the substrate before the onset of the splash \cite{Mani, Mandre}. On the other hand, the fact that, regardless of viscosity, splashes are invariably suppressed when the ambient pressure is sufficiently low suggests that there may be a common mechanism for both the violent corona and the slowly evolving thin sheet. It is therefore imperative that one investigate whether the splash mechanisms in these two cases have common features even though the timescales for corona (or thin-sheet) ejection and the overall shape of the splashing drops differ dramatically.  

This paper studies the onset of thin-sheet and corona ejection in the two cases. As previously noted, at high-$\mu$, thin-sheet ejection is delayed when the pressure is lowered \cite{Driscoll}. The major conclusion from the present work is that this is also true in the low-viscosity regime.  Corona ejection at low viscosity, which appears to occur immediately (that is, faster than the resolution of the cameras) at atmospheric pressure, is observed to be delayed at lower pressures.  Thus, prior to ejection of a thin sheet, the drop spreads as a thick fluid layer on the substrate.

The low-$\mu$ corona splash progresses through the same stages as a high-$\mu$ thin-sheet splash, albeit more rapidly: initial spreading is followed by thin-sheet ejection and eventual breakup of the thin sheet or corona. 
We find that the time of thin-sheet (or corona) ejection depends similarly on liquid viscosity and ambient gas pressure in both regimes. This argues in favor of a common mechanism for splashing.

In all experiments, we used silicone oils with dynamic viscosities $\mu$ ranging from 0.8 mPa$\thinspace$s to 19.0 mPa$\thinspace$s, with nearly constant surface tension $\sigma$ (between 17.0 dyn/cm and 20.6 dyn/cm), and density $\rho_L$ (between 0.76 g/cm$^3$ and 0.95 g/cm$^3$).
The low-viscosity results were duplicated using ethanol.
We generated drops of reproducible diameter $D$ of
$1.1\pm0.1$ mm, $1.8\pm0.1$ mm, or $3.2\pm0.1$ mm using nozzles of various sizes.
Drops were released from rest in a chamber from a height of 0.25 m to 1.0 m above dry smooth glass substrates (Fisher brand cover glass) to achieve an impact velocity $u_0$ from $2.0\pm0.05$ m/s to $4.0\pm0.1$ m/s.
A new substrate was used for each trial to avoid contamination from previous tests.
Ambient gas pressure $P$ in the chamber was varied from 5 kPa to 101 kPa.

Videos of drop impacts were captured from side or bottom views at up to 130,000 frames per second using a Phantom V12 or Phantom V1610 high speed camera.
Except for the lowest $\mu$ liquid, for which side views were needed, images from below were used to determine the time between impact and thin-sheet ejection $t_{ejt}$.
Images taken from the side were used to determine $D$ and $u_0$.

Figure \ref{images} shows the qualitative difference between a low-$\mu$ corona splash and a high-$\mu$ thin-sheet splash. 
Xu \cite{Xu2} demonstrated that there is a non-monotonic dependence of threshold pressure on liquid viscosity. As shown in Fig. \ref{Pt}, $P_T$ initially decreases as $\mu$ is increased and then turns around and increases at high-$\mu$.  The minimum in this curve separates the low- and high- viscosity regimes of splashing.  
We note the existence of these two regimes is robust; the boundary between them is approximately constant at about 2 mPa$\thinspace$s over the explored parameter range of $D$ and $u_0$. 

\begin{figure}[h] 
\begin{center}
\includegraphics[width=2.4 in]{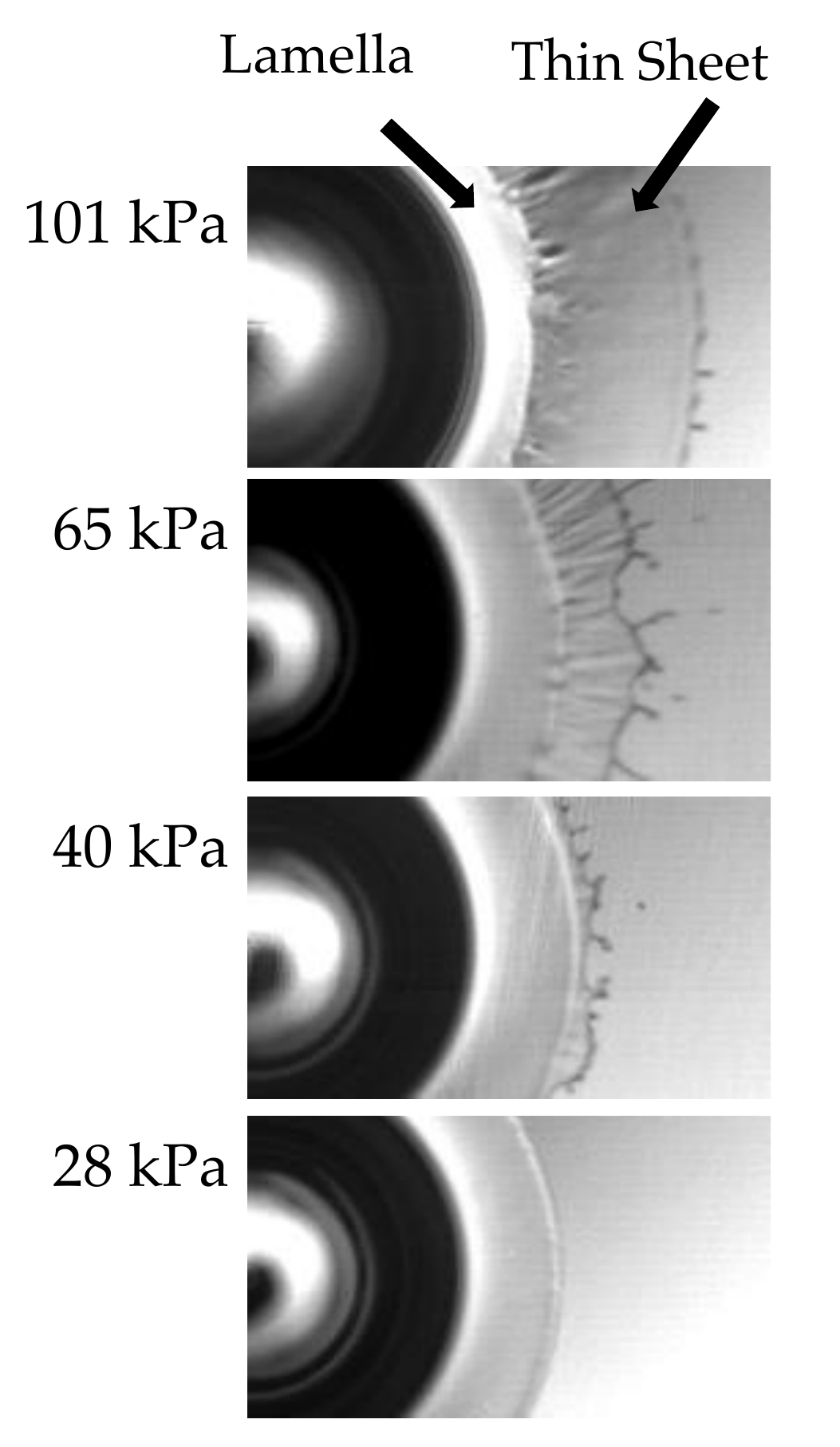}
\caption{Images taken from below of a 1.1 mPa$\thinspace$s silicone oil drop at 0.25 ms after impact. The splash occurs through the ejection of a thin liquid sheet from a thicker lamella. As $P$ is lowered, the sheet decreases in size until, at 28 kPa, no sheet is emitted.}
\label{p_images}
\end{center}
\end{figure}

Although there are two distinct viscosity regimes, the common dependence on ambient gas pressure hints at a connection between them. The first image of Fig. \ref{p_images} shows a bottom view of the corona splash at atmospheric pressure imaged at 0.25 ms after impact of a 1.1 mPa$\thinspace$s drop. This value of viscosity is well within the low-$\mu$ regime.
This image gives the true form of a low-$\mu$ splash: a thinner sheet, separate from the thicker lamella, breaks up into droplets. 
This structure has the same characteristics as that of a high-$\mu$ splash as described by Driscoll \textit{et. al.} \cite{Driscoll}. 
The subsequent images of Fig. \ref{p_images} show how the splash of such a low-$\mu$ drop evolves with pressure. 
As the ambient pressure is lowered, the ejected thin sheet is smaller, resulting in a smaller corona as previously observed \cite{Xu}. Below a pressure of 28 kPa there is no thin-sheet ejection and, as a result, there is no breakup or splash.  

\begin{figure}[h] 
\begin{center}
\includegraphics[width=3.1 in]{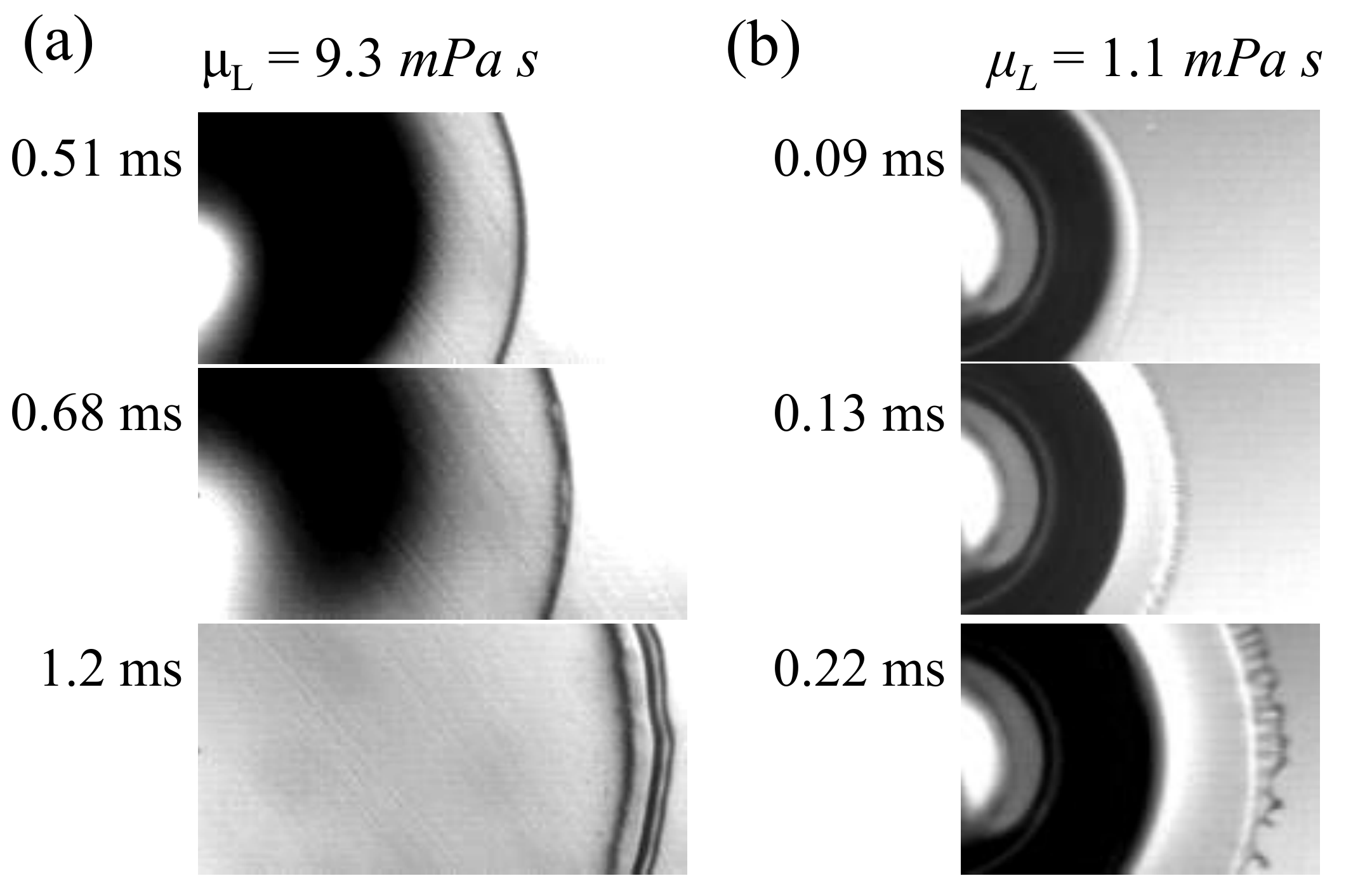}
\caption{Image time series of a 3.2 mm silicone oil drop after impacting a smooth surface at $u_0=4.0$ m/s and $P$=40 kPa. 
(a) The 9.3 mPa$\thinspace$s drop initially spreads on the surface forming a lamella. At 0.68 ms, a thin sheet first starts to be emitted above the surface. (b) The 1.1 mPa$\thinspace$s drop also spreads on the substrate before the sheet is ejected.}
\label{time}
\end{center}
\end{figure}

For high-$\mu$ liquids, there is a pronounced delay, $t_{ejt}$, between the moment of drop impact and the time of ejection of a thin sheet \cite{Driscoll}.  During this time, the drop spreads on the substrate as a thick film. In Fig. \ref{time}(a), we show the splash of a 9.3 mPa$\thinspace$s drop at $P$=40 kPa before, at, and after $t_{ejt}$. 
Initially, the drop spreads smoothly as if it will not splash.
However, at 0.68 ms a thin sheet is suddenly ejected from the advancing lamella; this sheet then grows, as seen in the third image of Fig. \ref{time}(a).
In Fig. \ref{time}(b), we demonstrate that the thin-sheet ejection of a low-$\mu$ splash is also not immediately following but is delayed. The ejection, however, occurs at only about 0.1 ms, which is much earlier than for high viscosity liquids. 

\begin{figure}[h] 
\begin{center}
\includegraphics[width=3.2 in]{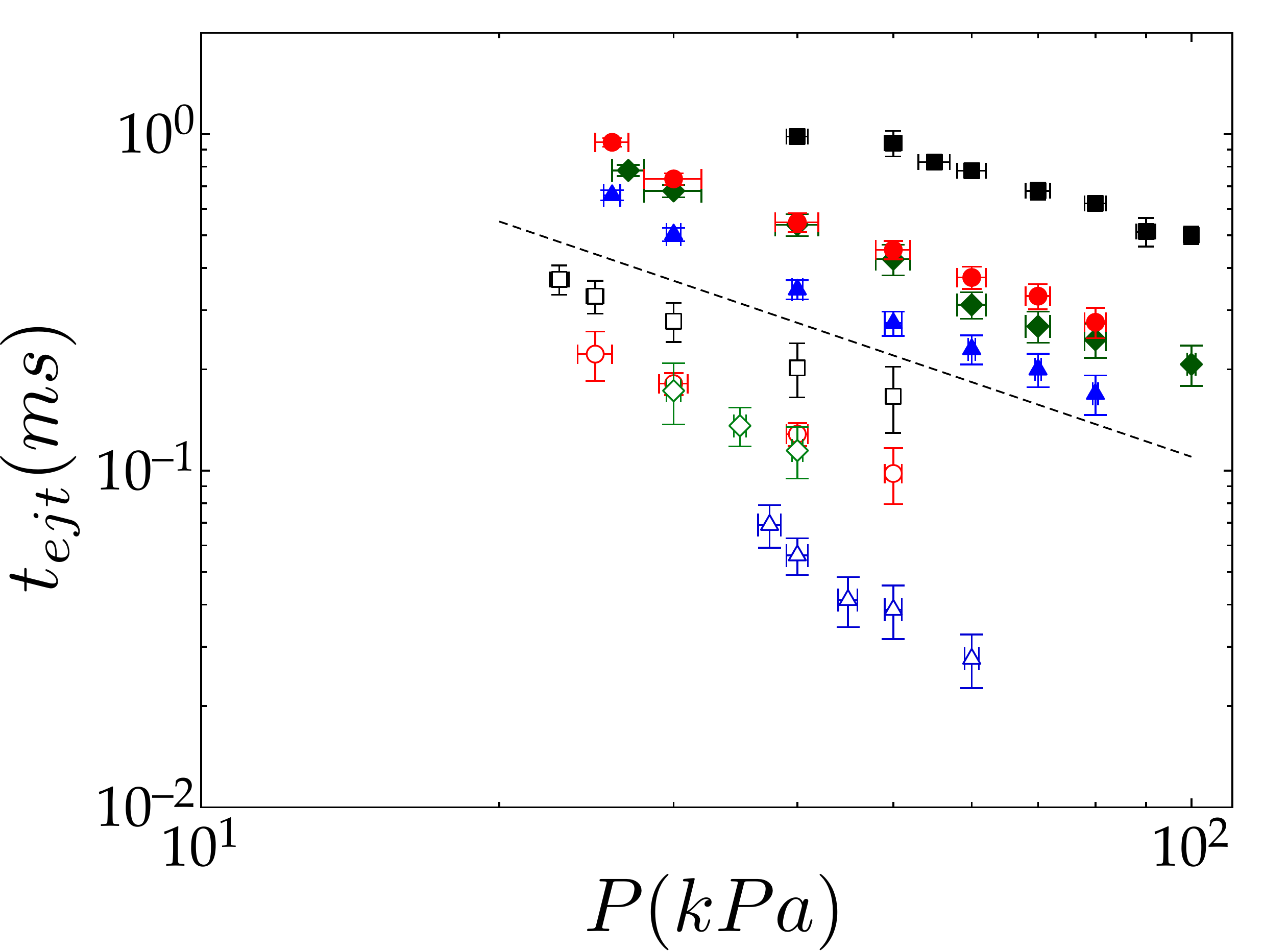}
\caption{Thin-sheet ejection time $t_{ejt}$ versus $P$ for $\mu$ = 0.8 ({\color{blue}
$\triangle$}), 1.1 ({\color{green}$\lozenge$}), 1.7 ({\color{red}$\circ$}), 2.3 ($\Box$), 4.2 ({\color{blue}
$\blacktriangle$}), 5.7 ({\color{green}$\blacklozenge$}), 
9.3 ({\color{red}$\bullet$}), and 19.0 ($\blacksquare$) mPa$\thinspace$s
 silicone oils. The impact velocity and drop size are fixed at $4.0\pm0.1$ m/s and $3.2 \pm 0.1$ mm, respectively. $t_{ejt}(P)$ increases with liquid viscosity. The dashed line serves as a guide to the eye between low-$\mu$ and high-$\mu$ regimes.}
\label{tejt}
\end{center}
\end{figure}

Figure \ref{tejt} shows $t_{ejt}$ as a function of gas pressure for both low and high viscosity drops.  Similar behavior is seen in both regimes.
Although $t_{ejt}$ is smaller for lower $\mu$ liquids, it still decreases approximately as $P^{-1}$ as it does for high-$\mu$ fluids \cite{Driscoll}.
Accordingly, near atmospheric pressure, the thin sheet is ejected at unresolveably low times for very low-$\mu$.  This explains why splashes of very low-$\mu$ liquids were mistakenly thought to occur immediately upon impact. However, we can observe sheet ejection of low-$\mu$ liquids because $t_{ejt}$ increases sufficiently as $P$ is lowered. 
As a result, we can resolve $t_{ejt}$ for liquids of $\mu$ as low as 0.8 mPa$\thinspace$s.

The non-monotonic viscosity dependence of threshold pressure indicates that there are two distinct regimes of liquid viscosity in splashing of drops on dry surfaces \cite{Xu2}. 
We have shown, however, that a splash evolves through thin-sheet formation for both regimes: 
the impacting drop first forms a thick lamella at the solid surface that initially spreads and only later ejects a thin liquid sheet. 
In the low-$\mu$ case, this thin sheet emerges at a large angle from the substrate to form the corona.  At high-$\mu$ the thin sheet, once formed, moves almost parallel to surface \cite{Driscoll2}.
The time of sheet ejection decreases with the ambient gas pressure in both regimes. 

Establishing that a delayed thin-sheet ejection is the common mechanism for splashing in both viscosity regimes entails several important consequences. (i) Any theory that relies on instantaneous splashing is precluded. (ii) Because it determines $t_{ejt}$, liquid viscosity is relevant even in the low-viscosity regime. (iii) Finally, focusing on the experimentally more accessible high-viscosity regime still provides important insight into low-viscosity splashing.

\begin{acknowledgments}
We are grateful to Irmgard Bischofberger and Michelle Driscoll for helpful discussions. This work was supported by NSF Grant DMR-1105145 and the facilities of the University of Chicago NSF-MRSEC, supported by DMR-0820054. C.S. acknowledges support from the NSF Graduate Research Fellowship Program.
\end{acknowledgments}

\bibliographystyle{apsrev4-1}
\bibliography{ThinSheetEjectionPaper.bib}
\end{document}